# A Multiwavelength View of ρ Oph I: Resolving the X-ray Source Between A and B

Sean J. Gunderson ,[1,2] Jackson Codd ,[3,2] Walter W. Golay ,[4,2] David P. Huenemoerder ,[1]

John M. Cannon ,[3,2] J. Alex Fluegel ,[5,2] Philip E. Griffin ,[6,2] Nathalie C. Haurberg ,[7,2]

Richard Ignace ,[8] Alexandrea Moreno ,[6,2] Pragati Pradhan ,[9] Alexis Riggs ,[7,2,10] James Wetzel ,[2,6,11]

Claude R. Canizares ,[1] and the MACRO consortium[2,*]

[1] Kavli Institute for Astrophysics and Space Research, Massachusetts Institute of Technology , 77 Massachusetts Ave., Cambridge, MA 02139, USA

[2] The MACRO Consortium; http://www.macroconsortium.org

[3] Department of Physics and Astronomy, Macalester College, 1600 Grand Avenue, Saint Paul, MN 55105, USA

[4] Center for Astrophysics | Harvard & Smithsonian, 60 Garden St., Cambridge, MA 02138, USA

[5] MakerSpace Network, Knox College, 2 E. South Street, Galesburg, IL, 61401, USA

[6] Department of Physics and Astronomy, University of Iowa, 30 N. Dubuque Street, Iowa City, IA 52242, USA

[7] Department of Physics and Astronomy, Knox College, 2 E. South Street, Galesburg, IL, 61401, USA

[8] Department of Physics & Astronomy, East Tennessee State University, Johnson City, TN, 37614, USA

[9] Department of Physics and Astronomy, Embry-Riddle Aeronautical University 3700 Willow Creek Road Prescott, AZ, 86301

[10] Minnesota Institute for Astrophysics, University of Minnesota, 116 Church St. SE Minneapolis, MN 55455

[11] Department of Physics, Coe College, 1220 1st Ave NE, Cedar Rapids, IA 52402

## ABSTRACT

We present a multiwavelength analysis of the central stellar pair of ρ Oph, components A and B. Using recent high-resolution *Chandra X-ray Observatory* observations, we demonstrate with high confidence that the dominant X-ray source is ρ Oph B, while ρ Oph A is comparatively X-ray faint. This result contrasts with earlier *XMM-Newton* observations, which, due to limited spatial resolutions, attributed the X-ray emission to ρ Oph A. An analysis of ρ Oph B's X-ray light curves and spectra reveals properties more consistent with a cool star than a hot star. We therefore propose that ρ Oph B is an Algol-like binary system, consisting of a B-type primary and an active, X-ray-emitting GK-type companion.

## 1. INTRODUCTION

As the namesake of the ρ Oph molecular cloud complex, ρ Oph consists of several young, massive stars spanning early formation to the main sequence. The primary stars of the ρ Oph system, designated A, B, C, and DE, are massive stars of early to late B-types. Recent work has revealed new information about these stars, such as the magnetism in ρ Oph C (P. Leto et al. 2020).

In particular, ρ Oph A was determined to not only be X-ray bright by I. Pillitteri et al. (2017) from *XMM-*

Email: seang97@mit.edu

* A full list of MACRO Consortium authors for the current year is available at https://macroconsortium.org/about-us/members

*Newton* observations of the ρ Oph system but also a periodic source ($P \sim 1.205$ days). Recent optical observations have also discovered that ρ Oph A is a binary system (M. E. Shultz et al. 2025) of two B-type stars, hosting a well-ordered dipolar magnetic field (R. Klement et al. 2025). A list of stellar properties for ρ Oph A's system is given in table 1 along with the known properties of ρ Oph B.

One of the proposed mechanisms by I. Pillitteri et al. (2017) for the periodic X-ray flux was a misaligned magnetic field causing a hotspot on the surface of the star, so the analysis by I. Pillitteri et al. (2017) appears robust. It should be noted, however, that the astrometric determination of ρ Oph A as the X-ray source was made using the *XMM* Metal Oxide Semiconductor (MOS) 1 detector. This detector has a FWHM PSF of 4.4" whereas ρ Oph A and B are separated by only 3" (M. A. Cordiner

**Table 1.** Stellar Properties of $\rho$ Oph AB

| Star | Parameter | Value | Reference |
|------|-----------|-------|-----------|
| A | Spectral Type Aa | B2/3 V | (a) |
| | Spectral Type Ab | B2/3 V | (a) |
| | $M_{\mathrm{Aa}}$ ($M_\odot$) | 8.4 | (a) |
| | $M_{\mathrm{Ab}}$ ($M_\odot$) | 6.0 | (a) |
| | $R_{\mathrm{Aa}}$ ($R_\odot$) | 4.2 | (a) |
| | $R_{\mathrm{Ab}}$ ($R_\odot$) | 3.1 | (a) |
| | $T_{\mathrm{Aa}}$ (kK) | 23 | (a) |
| | $T_{\mathrm{Ab}}$ (kK) | 19 | (a) |
| | $\log(L_{\mathrm{Aa}}/L_\odot)$ | 3.63 | (a) |
| | $\log(L_{\mathrm{Ab}}/L_\odot)$ | 3.02 | (a) |
| | $B_{\mathrm{Ab},0}$ (kG) | 0.2 | (a) |
| | $B_{\mathrm{Ab},1}$ (kG) | 0.98 | (a) |
| | $P_{\mathrm{Ab,rot}}$ (d) | 0.747326 | (a) |
| | $P_{\mathrm{orbit}}$ (d) | 88.00 | (a) |
| | $a$ (AU) | 1.1 | (a) |
| | Age (Myr) | 15 | (a) |
| B | Spectral Type | B2 V | (b) |
| | $M_{\mathrm{B}}$ | 8 | (c) |

Note—(a) M. E. Shultz et al. (2025); (b) N. Houk & M. Smith-Moore (1988); (c) C. Allen et al. (2018)

et al. 2013). Thus, while source positioning can rule out one of the stars based on the shape of the event's centroid, this requires precise (relative) astrometric corrections to a detector's aspect solution.

Alternatively, a more precise instrument can be used to resolve the two stars; a feat that only the *Chandra X-ray Observatory* can do in X-ray. Here we report on such an observation that made use of *Chandra*'s High Resolution Transmission Gratings (HETG) to acquire high-resolution spectra and more precise astrometry. We also report on data from optical and radio observations to further investigate the systems' properties. This paper is the first in a broader effort by the MACRO consortium to conduct a comprehensive, multiwavelength survey of the young stellar population in the $\rho$ Oph system.

This paper is organized as follows. In § 2, we provide details on the data we use and their reduction. In § 3, we perform source detections and relative position calibration using the high-resolution events from *Chandra* and optical images. In § 4, we analyze the temporal (4.1) and spectral (4.2) data on our sources. In § 5, we detail the radio data on our sources. Finally, in § 6, we give our conclusions.

## 2. DATA REDUCTION

A summary of the observations from public observatories used in our analysis is given in Table 2. The *Chandra* data was reprocessed with the standard pipeline in CIAO version 4.17 (A. Fruscione et al. 2006). *NuSTAR* data was processed with the standard pipeline in the HEASOFT version 6.34. *XMM* data was processed using SAS version 20.0.0 (C. Gabriel et al. 2004).

For each of the 10 *Chandra* observations, we made World Coordinate System (WCS) corrections to the event files using the optical coordinates of known stars in the field as reported in SIMBAD. The stars used are labeled in the top panel in Figure 1. WCS corrections we made using the `wcs_update` tool in CIAO. We then merged the *Chandra* observations after matching their celestial coordinate systems using a common tangent plane projection. The event file merging was done using `reproject_obs` in CIAO. We did not apply any WCS corrections to the *XMM* or *NuSTAR* files, nor did we merge them.

We also utilized data from the Robert L. Mutel Telescope (RLMT), owned and operated by the MACRO Consortium out of the Winer Observatory located in Sonoita, AZ, USA. The RLMT is a 0.51m PlaneWave CDK 20 telescope operated robotically using the `pyscope` software (W. Golay et al. 2024). After images were captured with the RLMT, WCS solutions were added using astrometry.net (D. Lang et al. 2010).

Finally, we analyzed archival radio observations of the $\rho$ Oph system from the Karl G. Jansky Very Large Array (VLA). $\rho$ Oph was observed in L-band (1-2 GHz) with the VLA in the most extended A-configuration (for which the characteristic synthesized beam is $\approx 1.3''$) with a 186 s on-source integration time. We reimaged the data using standard wide-field, high-sensitivity cleaning parameters using the CASA `tclean` routine ( CASA Team et al. 2022), including uniform weighting to minimize the size of the synthesized beam. The resulting sensitivity across the image was $\sigma_{\mathrm{RMS}} = 104\,\mu\mathrm{Jy}$.

## 3. SOURCE DETECTION

The first step of our analysis was to use the high spatial resolution of *Chandra* for an accurate astrometric solution. The central region of the merged *Chandra* event file is shown in the first panel of the lower portion of Figure 1. On top of this image, we placed 2" radius regions centered on $\rho$ Oph A and B's optical coordinates. $\rho$ Oph B is detected as the dominant X-ray source while $\rho$ Oph A is only weakly detected.

For a quality check on the positioning of these two stars, we compared the X-ray images to those collected with the RLMT. $\rho$ Oph was observed using the Sloan r' filter in 0.5-second exposures. One of these images is shown in the second panel of Figure 1 with the same regions as the *Chandra* image. The two stars are resolved by RLMT and match the optical coordinates.

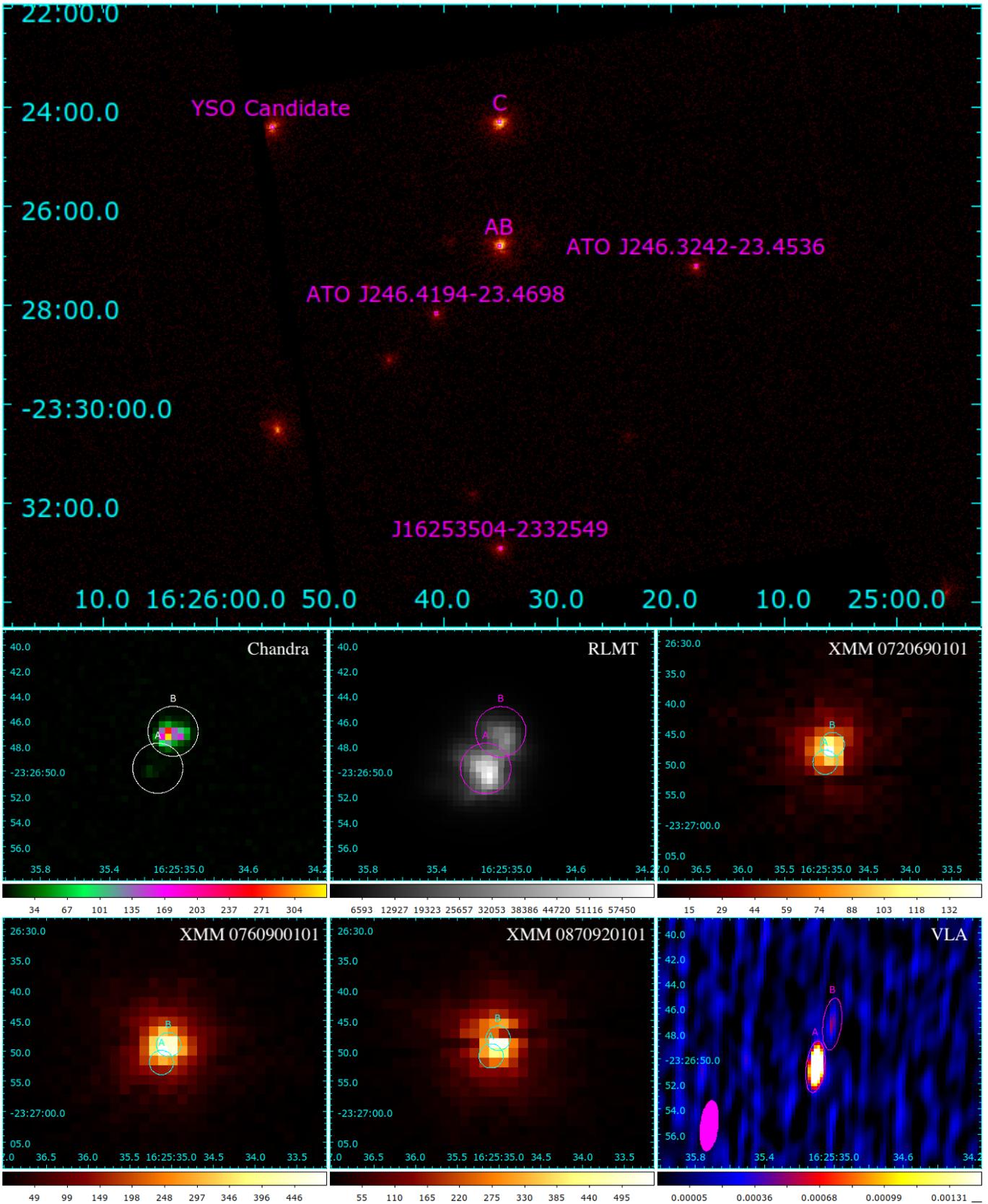

**Figure 1. Top panel:** *XMM* MOS1 event file from ObsID 0760900101 showing the reference stars used for correcting the aspect solutions in the X-ray observations. **Bottom panels:** Images from *Chandra* ACIS-S, RLMT, *XMM* MOS1 (in order of observation date), and VLA. Images are centered to the same WCS coordinate with north up and east to the left but have different scales. ρ Oph A's and B's positions are marked with 2" radius regions in each image, B being northwest of A. For *Chandra* and RLMT, the optical positions are used. For *XMM*, the presented regions are corrected source positions. Regions in the VLA use the uniform beam shape (shown as the solid ellipse) and have corrected source positions. See Table 3 for corrective values.

**Table 2.** $\rho$ Oph Observation Information.

| Observatory | Obs Id | Start Date | Exposure time (ks) |
|---|---|---|---|
| *XMM* | 0720690101 | 2013-08-29 | 53 |
| | 0760900101 | 2016-02-22 | 141.9 |
| | 0870920101 | 2020-09-13 14:26:50 | 79 |
| *NuSTAR* | 30601025002 | 2020-09-13 03:06:12 | 136.329 |
| VLA | 19A-446 | 2019-08-26 | 0.187 |
| *Chandra* | 24759 | 2022-03-21 | 18.18 |
| | 24760 | 2022-03-22 | 28.58 |
| | 26367 | 2022-03-26 | 9.93 |
| | 24763 | 2022-05-03 12:50:14 | 12.90 |
| | 26406 | 2022-05-03 22:09:19 | 16.84 |
| | 24672 | 2022-05-08 | 24.73 |
| | 26416 | 2022-05-12 | 24.53 |
| | 24761 | 2022-07-21 | 27.20 |
| | 24762 | 2022-07-24 | 10.92 |
| | 26481 | 2022-07-30 | 17.82 |

**Table 3.** Source position corrections.

| Obs Id | ΔRA ($''$) | ΔDec ($''$) |
|---|---|---|
| 0720690101 | -1.58 | -0.10 |
| 0760900101 | 0.00 | -2.14 |
| 0870920101 | -0.39 | -1.08 |
| 19A-446 | 0.89 | -0.85 |

The question now is the accuracy of the original determination of $\rho$ Oph A as the X-ray source by I. Pillitteri et al. (2017). In the three panels of Figure 1 labeled as *XMM*, we show the MOS1 event files. In these panels we show source regions that have been manually corrected to account for the aspect solution errors. For this, we started with the known optical position of each of the reference stars used for *Chandra* WCS corrections. We then applied a common shift to the regions marking the sources' optical coordinates so that each region was centered on the centroid of the reference star events. The values of the shifts applied are given in Table 3.

Each *XMM* observation had RA and Dec offsets of order 2".[12] In particular, the aspect solution of observation 0760900101 was off by 2.14" in declination. Without correcting this, it appears as though A is the X-ray source. With this correction, B's position more squarely falls on the brightest portion of events. From the other two *XMM* observations, however, the source of the X-rays is less certain. This is due to the 4.4" FWHM PSF of the MOS1 detector. Given that $\rho$ Oph A and B are 3" in separation, it is unreliable to use just *XMM* for source detection. It is only with the *Chandra* observations that we can determine which of the two sources is the dominant X-ray source. Taken together, these observations provide compelling evidence that $\rho$ Oph B is the primary X-ray source.

## 4. X-RAY OBSERVATIONS

### 4.1. *Temporal Analysis*

In Figure 2, the concatenated *Chandra* first-order grating light curves with 2 ks bins are displayed in the top panel, and the *NuSTAR* and *XMM* EPIC-pn light curves with 1 ks bins are shown in the bottom panels.

One of the proposed explanations for the variability in the *XMM* data by I. Pillitteri et al. (2017) was a rotating, offset magnetic field, which Obs Id 0760900101 supported when phased to a 1.205 day period. However, the simultaneous *XMM* and *NuSTAR* observation captured an X-ray enhancement event that is morphologically similar to a cool star flare (e.g., the many flares from EV Lac, D. P. Huenemoerder et al. 2010). In the bottom left panel of Figure 2, we show the light curve of this event for the *NuSTAR* FPMA (black) and *XMM* EPIC-pn (magenta) instruments, plotted on the same reference time of the start of the *NuSTAR* observation. While *XMM* did not capture the entire length of the flare, *XMM* did observe the first reheating event ($\sim$ 50 ks) at the same time as *NuSTAR* along with a second reheating ($\sim$ 80 ks).

These individual events will be discussed in more detail in § 4.2, but here we can make qualitative descriptions. Based on the differences in effective area and sensitivity between *XMM* and *NuSTAR*, the reheating events were not significant in strength. The soft response of *XMM* allowed it to see the first reheating event with much more detail and to completely observe the second. We can determine from the EPIC-pn light curve that these reheating events extended the lifetime of the hot plasma from the flare.

We next phased these light curves to the 1.205 day period from I. Pillitteri et al. (2017), using the start of Obs Id 0720690101 as phase $\phi = 0$. The resulting plot is shown in the bottom right panel of Figure 2. Unlike the enhancement event studied by I. Pillitteri et al. (2017) in Obs ID 0760900101, the flare-like event observed by *NuSTAR* did not occur at the same phase. We can use this to rule out the possibility of a repeating hot spot from a misaligned magnetic field as the source of the X-rays and variability. Taking into account the morphology of the event, we argue that this is evidence that $\rho$ Oph B has a cool star companion.

We would expect flares to occur frequently if a cool star were present in the system. The *Chandra* observations can provide an estimate of this frequency, as it spanned approximately 4 months with 1-month gaps between each portion. During this time, 2 potential flares, Obs Id's 26416 and 26481, were captured, giving a lower limit flare rate of >0.017 day$^{-1}$. This is on the lower end of the flare rates of young GKM-type stars but is within nominal rates, especially for G stars (A. D. Feinstein et al. 2024).

These properties suggest that $\rho$ Oph B is similar to the star Algol, which also consists of a B-type star with a cooler K-type companion (J.-F. Lestrade et al. 1993). In Algol, the B star is the dominant source of visible ra-diation, but at X-ray wavelengths, the coronal emission of the K star dominates (P. Testa et al. 2007). In the radio, large coronal loops track the cooler K-type star through the entire orbital period (W. M. Peterson et al. 2010), which has been observed in at least one other Algol-analog (HR 1099, W. W. Golay et al. 2024).

---

[12] This is expected behavior of the EPIC system, based on the observatory guide.

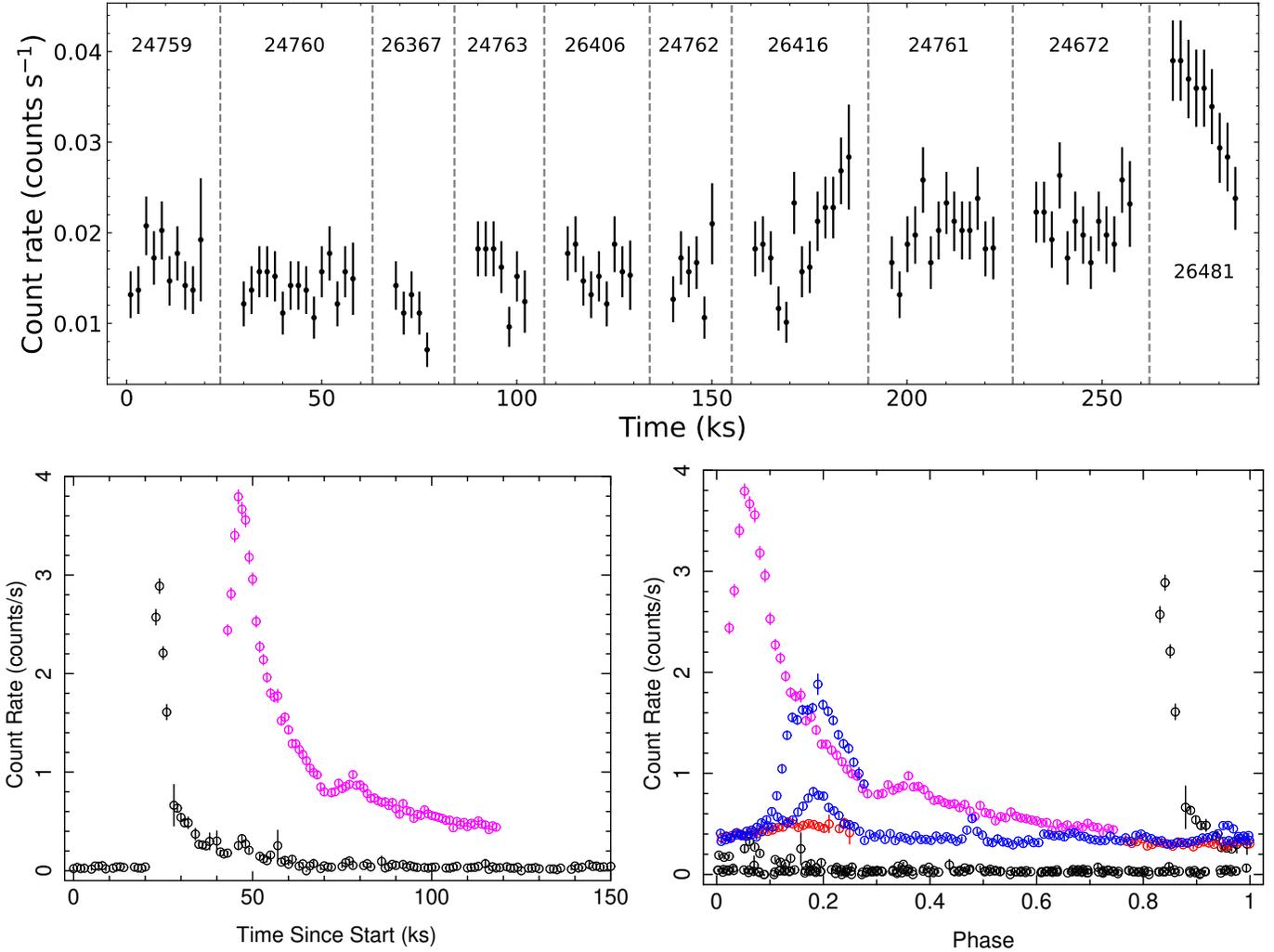

**Figure 2. Top:** Concatenated *Chandra* light curve in observation order with 2 ks time bins. **Bottom left:** *NuSTAR* FPMA (black) and *XMM* PN (magenta; Obs Id 0870920101) light curves with 1 ks bins. **Bottom right:** *NuSTAR* FPMA (black) and *XMM* PN (red; Obs Id 0720690101, blue; Obs Id 0760900101, and magenta; Obs Id 0870920101) light curves with 1 ks bins phased to the beginning of Obs Id 0720690101 assuming a 1.205 day period.

It is worth noting that we are not suggesting that $\rho$ Oph B is an evolved, semi-interacting system like Algol. The stars in $\rho$ Oph are very young, on the order of 15 Myr based on the age of $\rho$ Oph A (M. E. Shultz et al. 2025). Algol on the other hand is an evolved system of approximately 500 Myr. Our comparison is limited only to Algol as a (possibly) prototypical hot+cool star binary system. Such systems are rare and likely with few members to provide sufficient understanding of what their main-sequence properties are like.

### 4.2. *Spectral Analysis*

To determine whether the X-ray data of $\rho$ Oph B is consistent with a cool star, we analyzed the spectra from the observations using thermal plasma models. Specifically, we conducted three model fits based on the temporal features: Quiescence, Peak Flare, and the primary reheating event. Model fits were done using the Interactive Spectral Interpretation System (ISIS; J. C. Houck & L. A. Denicola 2000). We assumed Poisson statistics and used a Cash fit statistic (W. Cash 1979). The table of best-fit parameters is given in Table 4.

For the quiescent state, we utilized the *Chandra* High Energy Grating (HEG) and Medium Energy Grating (MEG) and the *XMM* Reflection Grating Spectrometer (RGS) from Obs ID 0720690101. These datasets were fitted with a two-temperature apec model (A. R.

**Table 4.** Plasma model best fit parameters with 68 percent errors

| Parameters | Value | | | Unit |
|---|---|---|---|---|
| | Quiescence | Peak Flare | Reheating | |
| $T_1$ | $8.80^{+0.48}_{-0.59}$ | – | $9.20^{+0.17}_{-0.20}$ | MK |
| $T_2$ | $28.01^{+2.48}_{-1.60}$ | $61.63^{+2.32}_{-2.16}$ | $32.74^{+2.43}_{-0.48}$ | MK |
| $Norm_1$ | $6.99^{+1.58}_{-1.13}$ | – | $7.91^{+0.94}_{-0.40}$ | $10^{-18}$ VEM/$4\pi d^2$ |
| $Norm_2$ | $12.69^{+0.79}_{-0.85}$ | $30.83^{+0.97}_{-0.95}$ | $33.13^{+0.63}_{-0.51}$ | $10^{-18}$ VEM/$4\pi d^2$ |
| $A$(O) | $0.21^{+0.07}_{-0.06}$ | – | $0.88^{+0.08}_{-0.15}$ | Relative to solar |
| $A$(Ne) | $1.02^{+0.2}_{-0.2}$ | – | $2.37^{+0.10}_{-0.26}$ | |
| $A$(Mg) | $0.16^{+0.05}_{-0.04}$ | – | $0.56^{+0.07}_{-0.08}$ | |
| $A$(Si) | $0.22^{+0.05}_{-0.04}$ | – | $0.29^{+0.09}_{-0.05}$ | |
| $A$(S) | $0.10^{+0.10}_{-0.06}$ | – | $0.31^{+0.10}_{-0.06}$ | |
| $A$(Fe) | $0.08^{+0.03}_{-0.02}$ | $0.27^{+0.04}_{-0.04}$ | $0.32^{+0.02}_{-0.02}$ | |
| $N_{\rm H}$ | 0.14 | – | $0.37^{+0.01}_{-0.01}$ | $10^{22}$ cm$^{-2}$ |
| $\mathcal{F}_X$ | 1.18 | 4.87 | 3.80 | $10^{-12}$ ergs cm$^{-2}$ s$^{-1}$ |
| $L_X$ | 2.70 | 11.13 | 8.68 | $10^{30}$ ergs s$^{-1}$ |
| $\mathcal{F}_R$ | 0.23 | 0.94 | 0.73 | mJy |

Note—Quiescence used all *Chandra* observations and *XMM* observation 07260900101. Peak Flare used the *NuSTAR* observation. Reheating used *XMM* observation 0870920101. Solar Abundances are from E. Anders & N. Grevesse (1989). The absorption used for the quiescent state was frozen. Reported X-ray fluxes are unabsorbed in the 0.2–12 keV band. X-ray luminosities assume a distance of $d = 138.2$ pc ( Gaia Collaboration et al. 2021). Radio fluxes calculated assuming a Güdel-Benz relation for Algol systems at 5 GHz (M. Guedel & A. O. Benz 1993; A. O. Benz & M. Guedel 1994).

Foster et al. 2012) with an interstellar absorption component `tbabs`. The interstellar Hydrogen column was fixed to a constant value of $N_H = 0.14$ ( HI4PI Collaboration et al. 2016). We found that this two-temperature plasma model accurately replicated the spectrum, which is shown in the top panel of Figure 3. Note that because of the degradation of *Chandra*'s soft response, we truncated the *Chandra* spectra to 10 Å. The RGS was also truncated to 22 Å due to a high background occurring for this observation.

The spectral features of the quiescent emission are more similar to cool stars than B stars. Emission lines are narrow, consistent with being unresolved, having an upper limit of about 600 km s$^{-1}$ FWHM, as best constrained by the Mg XII Lyman-$\alpha$-like line in the HETG spectrum. The 90-percent contours of the line widths are shown in Figure 4 as the solid line. For comparison, the similarly sharp, unresolved line widths of the coronal source HR 1099 (K1 IV + G5 V + K3 V) are shown as the dashed 90-percent contours ( D. P. Huenemoerder et al. 2013).

The $\rho$ Oph B spectra also show evidence of an inverse first-ionization potential (FIP) effect in lines from the strong Ne X H-like Ly-$\alpha$ line and weak Fe emission ( D. P. Huenemoerder et al. 2013). Our plasma fitting shows an overabundance of Ne compared to other metals, which is also consistent with a FIP effect. An OB star's X-ray spectra, on the other hand, show stronger emission lines from all metals with a much weaker continuum and broadened lines with at least 1000 km s$^{-1}$ width ( P. Pradhan et al. 2023). Additionally, OB stars are poorly described by plasma models of only a few temperatures, instead requiring more detailed differential emission measure distributions of at least 6 temperatures ( D. P. Huenemoerder et al. 2020; D. H. Cohen et al. 2021).

While a cool star explains the emission overall, there are features at 13 Å and between 16 − 17 Å that are difficult to explain. The RGS has several chip gaps and bad pixels that can cause a loss of signal in different spectral bins. However, the observed features are uncharacteristically broad to be caused by these detector faults. Alternatively, there are a significant number of

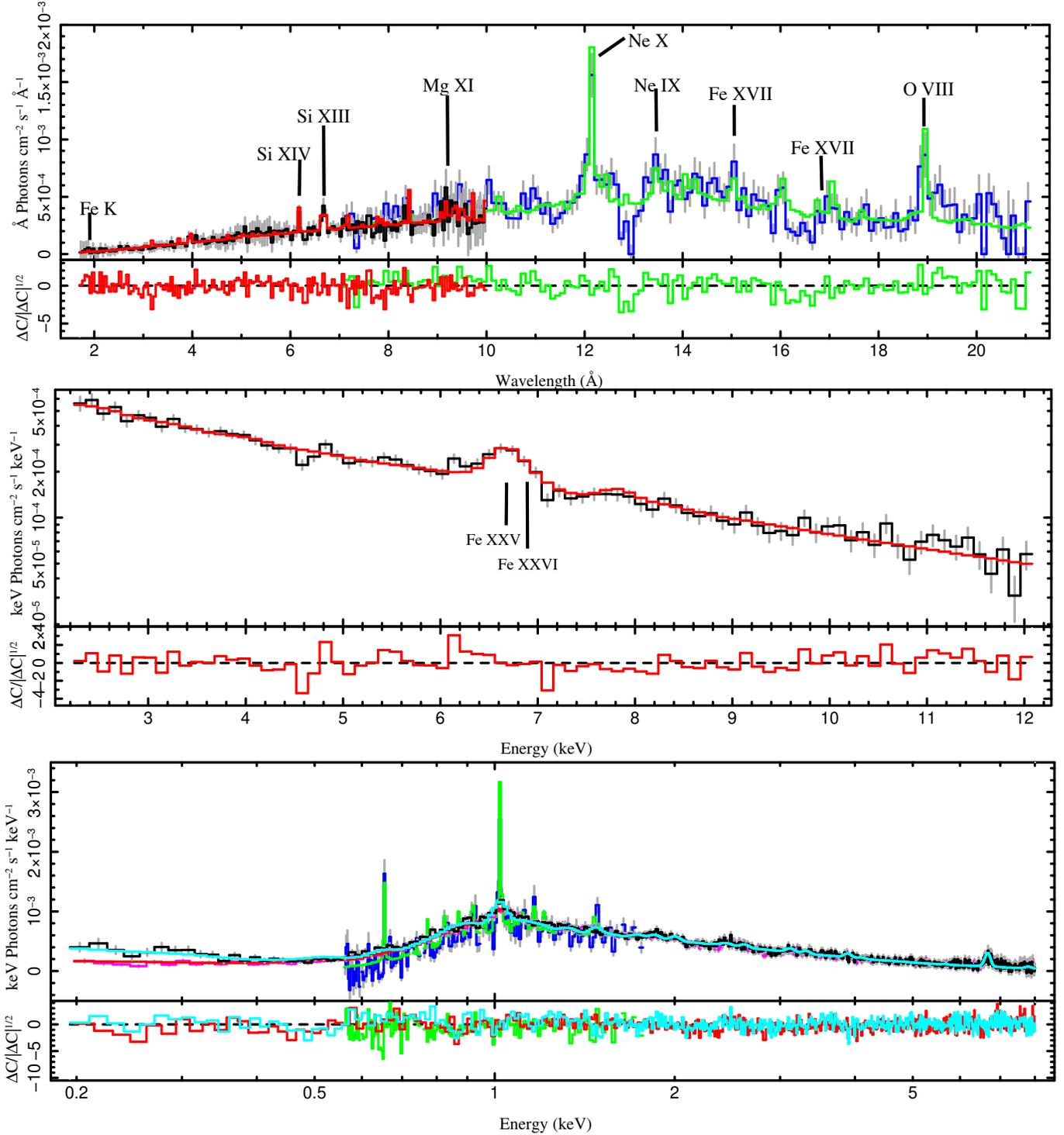

**Figure 3. Top panel:** total *Chandra* HEG+MEG spectrum (black) and RGS1+RGS2 (blue) with best fitting two-temperature plasma model (red and green) during quiescence. Both spectra have been binned by a constant factor of ten. **Middle panel:** *NuSTAR* FPMA+FPMB (black) with best fitting one-temperature plasma model (red), binned by a constant factor of three. **Bottom panel:** *XMM* PN (magenta), MOS1+MOS2 (black), and RGS1+RGS2 (blue) with best fitting two-temperature plasma model (red, cyan, and green, respectively). The PN and MOS spectra are binned by a constant of three while RGS is binned by ten.

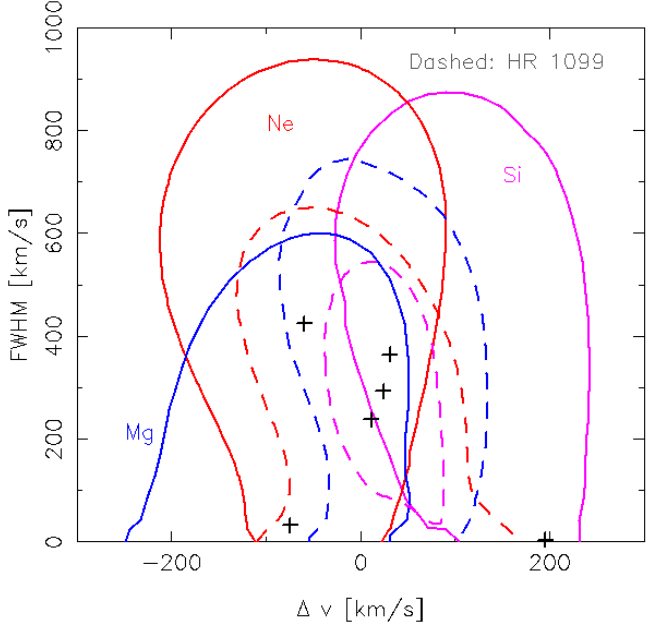

**Figure 4.** Ninety percent line width contours of $\rho$ Oph (solid) compared to the known unresolved line widths of HR 1099 (dashed).

Fe ion lines in both these regions, so absorption features could explain the spectrum dips. The chief problem is explaining why such highly ionized Fe is not in emission.

We next focused on the main flare peak observed by *NuSTAR*, shown in the middle panel of Figure 3. For this spectrum, we included the entire observation. Based on *NuSTAR's* high energy sensitivity and the high count rate during the flare, the main peak of the flare dominates the signal.

A single-temperature plasma model with no absorption component was used to fit the spectrum, with free parameters of temperature, normalization, and Fe abundance. Due to *NuSTAR's* spectral resolution and energy range, we could not constrain the abundances of other major elements. The 61.63 MK temperature of this model supports the cool-star flare interpretation since these temperatures are common in flares (J. P. Pye et al. 2015). Notably, the Fe abundance is three times larger than the quiescent emission. This is potentially a degeneracy effect of the change in temperature since line fluxes are determined by $A(\mathrm{El}) * VEM$. In other words, being at a lower-than-peak emissivity temperature, the elemental abundances can be artificially enhanced in the model fitting. For the Fe K region, the 61.63 MK is near the peak of the emissivity curve for Fe XXV He-like Lyman-$\alpha$ (66.83 MK), so the almost three times increase in line normalization suggests that a genuine abundance change cannot be ruled out.

For the final spectra, we used the PN, MOS, and RGS detectors from *XMM* to analyze the reheating event in comparison to the main peak. These spectra are plotted in the bottom panel of Figure 3. Here, we again used a two-temperature plasma model, as in the quiescent case; however, this time we found that the absorption could not be frozen to the nominal ISM value. During the reheating event, the source spectrum experienced enhanced intrinsic absorption. In this flare interpretation, this increased absorption could be the result of ejected material (C. Argiroffi et al. 2019; A. M. Veronig et al. 2021; H. Chen et al. 2022), but such material would need to be far denser than the cool star's corona. Such could be the case if X-ray emission originated from magnetic loop foot-points viewed through an overlying arcade of loop-confined plasma (K. Bicz et al. 2024).

The metal abundances also suggest a violent ejection of material that stretched down to the chromosphere. All freely fitting elements, except Si, show significant increases in their abundances that are statistically different from the quiescent spectra with high confidence. This suggests that the detected flare was substantial enough in strength to dredge up material from the chromosphere, populating the corona with more metals. A similar effect has been observed in known cool-star flaring system, e.g., Algol, HR 1099, and CN Leo (R. Nordon & E. Behar 2008; C. Liefke et al. 2010).

If the X-ray source in $\rho$ Oph B is a B star, then we would expect its X-ray luminosity to follow the usual $L_X/L_{\mathrm{Bol}}$ trend of B stars (e.g., D. H. Cohen 2000), and have $\log(L_X/L_{\mathrm{Bol}}) \sim -8$. Using the mass-luminosity relationship for main-sequence stars

$$\log\left(\frac{L_{\mathrm{Bol}}}{L_\odot}\right) \approx 3.5 \log\left(\frac{M}{M_\odot}\right) \quad (1)$$

we can estimate $L_{\mathrm{Bol}}/L_\odot \approx 1448$ for the B-type star. This gives a quiescent X-ray luminosity ratio of $\log(L_X/L_{\mathrm{Bol}}) \sim -6.31$ while the peak flare reached $\log(L_X/L_{\mathrm{Bol}}) \sim -5.70$. These are orders of magnitude larger than expected for a normal B2 star. On the other hand, if the X-ray emitting source is a GK-type system, with mass of order $M_\odot$ so that $L_{\mathrm{Bol}} \approx L_\odot$, then these X-ray to bolometric luminosities are instead -3.15 in quiescence and -2.56 during the flare. These ratios are well within the expected upper limits for GK stars, providing additional support for our interpretation.

### 4.3. $\rho$ Oph A Zeroth Order Spectrum

Since $\rho$ Oph A and B are resolved in Chandra, we can also extract out the weaker flux from $\rho$ Oph A specifically. The flux from $\rho$ Oph A is too weak to have any measurable HETG grating arm on the detector. We instead ex-

**Table 5.** Power law DEM best fit parameters.

| Parameter | Value | Units |
|---|---|---|
| $\mathcal{F}_X$ | $0.037^{+0.005}_{-0.004}$ | $10^{-12}\,\mathrm{ergs\,cm^{-2}\,s^{-1}}$ |
| $\gamma$ | $1.22^{+3.78}_{-1.81}$ | |
| $T_{\max}$ | $26.61^{+13.38}_{-6.27}$ | MK |
| $N_{\mathrm{H}}$ | $0.58^{+0.35}_{-0.38}$ | $10^{22}\,\mathrm{cm^{-2}}$ |
| $L_X$ | $0.084^{+0.005}_{-0.004}$ | $10^{30}\,\mathrm{ergs\,s^{-1}}$ |

NOTE—Flux and luminosity are over $0.8 - 8\,\mathrm{keV}$. Upper limit on $\gamma$ is unconstrained.

tracted a zeroth order spectrum using 3 pixel radius regions centered on $\rho$ Oph A in each of the 10 *Chandra* observations. A background region with a 30 pixel radius was chosen nearby to not overlap with any other source. This gave us a total of 89 source counts and 3 background counts in the cumulative observation. The corresponding spectrum is shown in Figure [ADD LATER].

We fit the spectrum with an absorbed power law differential emission measure (DEM) distribution

$$\frac{d\mathrm{EM}}{dT} = D_0(d)\left(\frac{T}{T_{\max}}\right)^{\gamma},\qquad(2)$$

where $T_{\max}$ is the maximum temperature of the plasma, $\gamma$ is the slope of the distribution, and

$$D_0(d) = (\gamma+1)\frac{T_{\max}^{\gamma}}{T_{\max}^{\gamma+1} - T_{\min}^{\gamma-1}}\frac{\mathrm{EM}}{4\pi d^2},\qquad(3)$$

and EM is the volume emission measure. This power law DEM is constructed through a weighted sum of isothermal plasma models; for more details on the model, see D. P. Huenemoerder et al. (2020).

Due to the limited counts in the spectrum, constraining EM introduces significant errors. Specifically, the problem is in constraining the slope $\gamma$. The upper limit of $\gamma$ was unconstrained; it always ran up to the maximum limit we placed on it. We accounted for this by limiting the search window to an upper limit of $\gamma = 5$ and used the model flux as a free parameter through the `cflux` model component. Additionally, since we are only have a zeroth order spectrum, we do not have any resolved emission lines to constrain the elemental abundances. All abundances were frozen to unity, relative to solar from E. Anders & N. Grevesse (1989). The free parameters were the flux, $\gamma$, $T_{\max}$, and $N_{\mathrm{H}}$. The best fit model parameters are given in Table 5.

While we could not constrain the upper limit on $\gamma$, the lower limit was well constrained and notable at $-0.59$. As a point of comparison, the prototypical embedded wind shock star $\zeta$ Pup (O4 I) has a DEM with a slope of $\gamma = -2.6$ (D. P. Huenemoerder et al. 2020). The difference in slope is not due to the spectral region limitations since D. P. Huenemoerder et al. (2020) conducted their analysis of $\zeta$ Pup on recent data that was also limited to photons above 1 keV. We can infer from the shallower, potentially positive, slope that there is significantly more hot plasma relative to cool in $\rho$ Oph A than in $\zeta$ Pup.

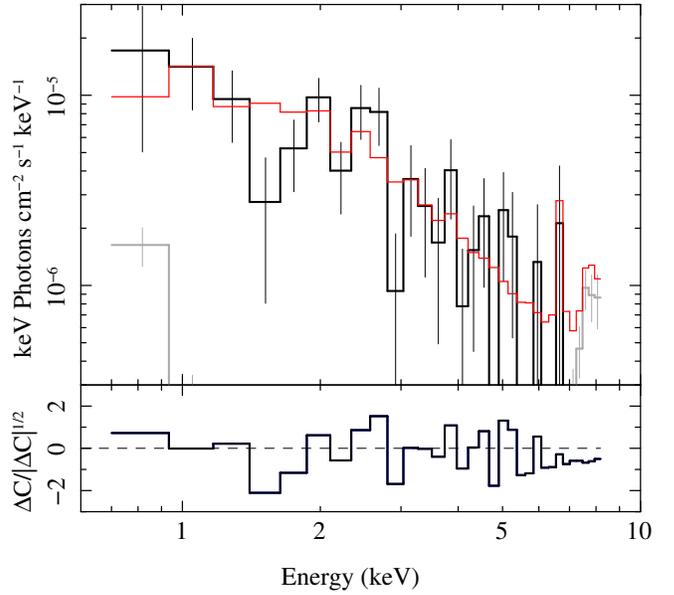

**Figure 5.** Zeroth order *Chandra* ACIS-S spectrum of $\rho$ Oph A (black) with best fitting power law DEM model (red). The spectrum of the background region is shown in gray.

The comparison against $\zeta$ Pup grows more interesting when considering the maximum temperature, which for $\zeta$ Pup is only approximately $T_{\max} = 12$ MK. In the embedded wind shock framework, O stars have a higher maximum temperature due to their faster wind terminal speeds and denser winds (P. Pradhan et al. 2023). Since $\rho$ Oph A's maximum temperature is higher than $\zeta$ Pup's despite being a later spectral type, it is unlikely to be producing X-rays through embedded wind shocks.

The magnetic field in the system can explain the differences in the DEM. Magnetically confined wind shocks can produce much hotter plasma (Y. Nazé et al. 2014). Assuming the majority of the X-rays are coming from $\rho$ Oph Ab's magnetic field, the X-ray luminosity scales as $\log(L_X/L_{\mathrm{Bol}}) \approx -7.68$. This is on the higher end of expect B2/3 star X-ray luminosities (D. H. Cohen 2000), but it is consistent with magnetic systems being generally brighter. Testing this hypothesis for magnetic confinement, however, will take a substantial monitor-ing campaign with *Chandra* to get sufficient counts in *Chandra*'s dispersed gratings.

## 5. RADIO ANALYSIS

If $\rho$ Oph B contains a cool star, then we can further predict what its radio flux is from its X-ray data. Assuming it is an Algol-like system, then the Güdel-Benz relation (M. Guedel & A. O. Benz 1993; A. O. Benz & M. Guedel 1994) for the system will be

$$\frac{L_R}{L_X} \approx 1.92 \times 10^{-15} \ \text{Hz}^{-1} \qquad (4)$$

at a frequency of 5 GHz. The quiescent X-ray luminosity would then correspond to a radio flux of $\mathcal{F}_R = 0.23$ mJy. Similar calculations are given for the flare and reheating events at the end of Table 4.

This is well detectable by the VLA, so we reduced the archival L-band (1–2 GHz) data of the $\rho$ Oph system for a final source detection. This observation is shown in the lower right panel of Figure 1 for a uniformly weighted field. Source regions for $\rho$ Oph A and B are based on the beam size and orientation of the observation, shown as the solid ellipse.

$\rho$ Oph A is clearly detected, as is expected for a magnetic system (R. Klement et al. 2025). $\rho$ Oph B is also detected with a flux density of $0.402 \pm 0.107$ mJy at a $4\sigma$ significance. If we take this ~1.5 GHz measurement and our predicted 5 GHz flux density, it would correspond to a power law radio SED with spectral index $\alpha = -0.46 \pm 0.22$. Such a spectral index is well within reason for a GK star (e.g., Algol, HR 1099, & UX Arietis, W. W. Golay et al. 2023).

## 6. CONCLUSIONS

We reported on a multi-wavelength analysis of the AB component of the $\rho$ Oph system. Prior works have attributed the X-ray emission to $\rho$ Oph A; however, recent *Chandra* observations have revealed that the X-ray source is more accurately localized to $\rho$ Oph B. Moreover, the *Chandra* data revealed that $\rho$ Oph A is relatively faint in X-rays.

In comparison, $\rho$ Oph B's brightness is significant and constant, even in quiescence. From the *Chandra* HETG and *XMM* RGS data, the flux in quiescence is $1.18 \times 10^{-12}$ ergs cm$^{-2}$ s$^{-1}$. Such a high flux for this system is difficult to explain as coming from a normal B2 main-sequence star. This is in addition to classical flare-like events in its light curves and spectral properties, which are more consistent with those of a cool star. These results are consistent with the presence of a GK-type companion in $\rho$ Oph B, though spectroscopic confirmation is still required.

As further evidence, the predicted and measured radio properties of the system, as determined from archival VLA data, also align with those of a cool star companion. This radio support is dependent on a single L-band (1 − 2 GHz) measurement, however, and does not fully characterize the radio SED.

This leads to two areas of future work. First is a deeper radio observation that samples the entire radio spectrum to determine the radio SED. We have planned for a series of VLA exposures across 1 − 12 GHz that will enable the high signal-to-noise constraint of the SED properties. Second, optical spectroscopic observations should be taken to measure a radial velocity curve. While all of the discussed spectral properties align with a cool star companion, they are not definitive evidence of its presence. Only spectroscopic radial velocity observations will be able to confirm the existence of a companion star in $\rho$ Oph B and classify it as a "Demon Star."

## ACKNOWLEDGMENTS

Support for SJG and DPH was provided by NASA through the Smithsonian Astrophysical Observatory (SAO) contract SV3-73016 to MIT for Support of the Chandra X-Ray Center (CXC) and Science Instruments. CXC is operated by SAO for and on behalf of NASA under contract NAS8-03060.

This paper employs a list of Chandra datasets, obtained by the Chandra X-ray Observatory, contained in the Chandra Data Collection (CDC) 441 doi:10.25574/cdc.441

The National Radio Astronomy Observatory is a facility of the National Science Foundation operated under cooperative agreement by Associated Universities, Inc.

This research has made use of ISIS functions (ISISscripts) provided by ECAP/Remeis observatory and MIT (http://www.sternwarte.uni-erlangen.de/isis/).

The MACRO Consortium is an educational and research collaboration between Augustana College (Rock Island, IL), Coe College (Cedar Rapids, IA), Knox College (Galesburg, IL), Macalester College (Saint Paul, MN), and the University of Iowa (Iowa City, IA). MACRO operates the robotic 0.5m Robert L. Mutel Telescope (RLMT) at the Winer Observatory (Sonoita, AZ, USA) to foster immersive research and classroom experiences for undergraduate students. Data taken by MACRO is available upon request.

## AUTHOR CONTRIBUTIONS

SJG conducted the analysis of the X-ray data and was responsible for writing and submitting the manuscript.

DPH submitted the original Chandra observation proposal, validated the X-ray analysis, and edited the manuscript.

JC, JMC, and WWG conducted the radio analysis and edited the manuscript.

PEG, AM, and AR collected and calibrated the RLMT data.

All other authors provided theoretical and interpretative support along with reviewing the manuscript.